\documentclass{JHEP}
\preprint{
IMSc/2001/04/23\\
hep-th/0108002\\}
\title{
Interaction between two Fuzzy Spheres}
\author{Subrata Bal and Hiroyuki Takata \\
The Institute of Mathematical Sciences,\\
 CIT Campus,
Madras - 600 113,  INDIA \\
\email{subrata, takata@imsc.ernet.in}}

\abstract{We have calculated interactions between two fuzzy spheres
in 3 dimension.  It depends on the distance $r$ between the spheres 
and the radii $\rho_1$, $\rho_2$. 
There is no force between  the spheres when they
are far from each other 
(long distance case). We have also studied the interaction for $r=0$ case.
We find that an attractive force exists between two fuzzy sphere
surfaces.
}

\newcommand{\bel}{\begin{equation}\label}
\newcommand{\f}{\frac}
\newcommand{\non}{\nonumber \\}
\newcommand {\beq}{\begin{equation}}
\newcommand {\eeq}{\end{equation}}
\newcommand {\beqa}{\begin{eqnarray}}
\newcommand {\eeqa}{\end{eqnarray}}
\newcommand {\bc}{\begin{center}}
\newcommand {\ec}{\end{center}}

\def\a{\alpha}

\def\l{\lambda}
\def\vp{\varphi}
\def\e{\epsilon}
\def\s{\sigma}
\def\dag{\dagger}
\def\nn{\nonumber}

\def\vs5{\vspace*{5mm}}
\def\vs1{\vspace*{1cm}}
\def\vs2{\vspace*{2cm}}
\def\hs5{\vspace*{5mm}}
\def\hs1{\hspace*{1cm}}
\def\hs2{\hspace*{2cm}}
\def\vs50{\vspace*{50mm}}
\def\vs20{\vspace*{20mm}}

\begin{document}

\section{Introduction}

String theory in present form, 
include various types of extended objects 
like D-branes other than fundamental string. Therefore 
it is essential to reconstruct string theory such that it can treat 
them in unified way. The presence of D-branes 
brings the non-commutativity of space time. In 
fact we can derive non-commutative gauge theory on world 
volume of D-branes in string theory and that non-commutative 
gauge theory lead us to find corresponding  matrix model. 

IIB matrix model  is one of the proposals for 
non-perturbative reconstruction of string theory
\cite{ikkt1}.It is  a large $N$ reduced model of ten-dimensional 
supersymmetric Yang-Mills theory.
In matrix model the space time and matter dynamically emerge out. 
As the space time are represented by the diagonal elements
of the matrices the non-commutativity of space time is built in 
matrix model. At the same time, as the interactions are also 
described by the matrices  the non-commutative 
yang-mills theory in a flat back ground 
can be obtained by expanding the matrix model around a flat
non-commutative background \cite{noncmatrix}.
Such  non-commutative gauge  theory  
is obtained in string theory by introducing background constant B-field
\cite{noncstring}.
The non-commutative background is a D-brane-like background
which is a solution of the equation of motion. So, we can
study the D-brane in flat background within the framework 
of matrix model. We need to formulate a matrix model 
which gives us a scope to study the D-brane in 
curved background also.

Recently other different non-commutative backgrounds, for $e.g.$
a non-commutative sphere, or a fuzzy sphere
have also been studied \cite{iso1,isos}.
In \cite{iso1}, non-commutative gauge theories on fuzzy sphere
were obtained considering supersymmetric three dimensional matrix model
actions with a Chern Simon term, expanding around a classical solution. 
Although an ordinary matrix model has only
a flat background as a classical solution,
this matrix model can describe a curved background
owing to these terms.

Fuzzy sphere may correspond spherical D2-brane in string theory 
with background linear B-field in 
$S^3$ \cite{ars1,arss,hikida,koji,malda}. Specially $0$ radius one correspond to D0-brane. It is interesting 
to find out, in Matrix model, the object corresponding D2 or D0 brane in 
string theory. Also BPS objects corresponding BPS 
D branes are interesting.

In this paper, we will consider supersymmetric fuzzy sphere model 
in three dimension as in \cite{iso1}. We get this model by adding a Chern 
Simon term to the reduced model \cite{ikkt1}. We try to give a framework
which allows us to study a multi fuzzy sphere system.
We will study the two fuzzy spheres system in detail 
and try to investigate the interaction between  them.

This paper is organized as follows. In section 2, 
we present the model for the multi fuzzy 
sphere in background space. We calculate interaction of 
fuzzy spheres and space.
In section 3, 
we talk about the 
dynamics of the fuzzy spheres. We expand the action around a 
classical back ground and try to study the 
one loop interaction between fuzzy 
spheres in 
bosonic and supersymmetric case.
In particular, we have calculated the interaction between 
two fuzzy spheres.
We calculate the
potential for such system in both large and small 
distance case. This potentials are attractive for supersymmetric
case which  vanish for long distance.

\subsection{Space-time and brane from Matrix Model view point}

In general, in matrix model, we deal with  arbitrary Hermitian matrices. 
We can as well   artificially partition these matrices
into multiple blocks such that each diagonal block represents a part 
of space time (we call it an space time 
object or a brane.) and the off diagonal blocks represent the interaction 
between such branes. The size of such brane depends on the size of the 
matrix-block representing the brane. For example, block of size $1$,
describes a space time point.

Though the overall matrix is traceless, the individual blocks need not to be 
traceless and the value of trace of these blocks give the
space time co-ordinate of the center of the block (object).  
In this paper we assume that the trace belongs
to ${\bf R}^{10}$.

There may be some possibility of 'dynamical compactification' 
of 10 dimensional space time to $M^7 \otimes {\bf R}^3(S^3)$, which we are going
to assume  here \cite{ikktspace}.

It is known that there is non-commutative solution for classical equation of 
motion and non-commutative gauge theory on
such space time object (both plane and sphere case).Iso et al \cite{iso1}
 and others have shown that non-commutative gauge can be
realised on fuzzy sphere. For the flat 
case the gauge interaction can be
explained as the open strings which ends on the object (brane). 
The force between two of such branes at long
distance can be understood as close strings exchange between them. 
We like to understand whether this  
feature is valied for $M^7 \otimes {\bf R}^3(S^3)$ configuration of space time.  
In this paper we treat the
interaction of two fuzzy spheres in ${\bf R}^3$.

\section{Model for multi fuzzy spheres }

\subsection{Classical Picture}

We start with 
${\cal N}=2$ SUSY Yang-Mills-Chern-Simon reduced model

\bel{fuzaction}
 S= \frac{1}{g^{2}}Tr\Big(
-\frac{1}{4} [ A_{\mu} ,A_{\nu}] [A^{\mu} ,A^{\nu}]
  +\frac{2}{3}i\alpha \epsilon_{\mu \nu \lambda}
A^{\mu}A^{\nu}A^{\lambda}
  +\frac{1}{2}\bar{\psi}\sigma^{\mu}[A_{\mu},\psi]  \Big).
\eeq

\noindent
This is obtained by reducing the spacetime volume
of Yang-Mills-Chern-Simons theory to a single point \cite{iso1}
(c.f. Eguchi-Kawai and IKKT model \cite{ikkt1}).
The Chern-Simon term is added to the reduced model
to have fuzzy sphere solution as classical equations of motion.
$A_\mu$, $\psi_\alpha$ are $N\times N$ traceless Hermitian matrices.
$A_\mu$ is 3-dimensional vector and $\psi$ is two components Majorana spinor.
$\sigma_{\mu}(\mu=1,2,3)$ denote Pauli matrices. 
$\mu, \nu = 1\sim 3$, $\alpha, \beta = 1, 2$.
This action  is also obtained as  low energy effective action for 
spherical D2-brane in $S^3$,
using $SU(2)$ WZW model as string theory in $S^3$
\cite{ars1}. 

Action (\ref{fuzaction}) has $SO(3)$ global symmetry,   
$ A_{\mu} \rightarrow  A_{\mu}+ r_\mu {\bf 1}$ translation symmetry and
gauge symmetry by the unitary matrices
$
A_{\mu} \rightarrow UA_{\mu}U^{\dagger}\,\,, 
\hspace{0.3cm} \psi \rightarrow U\psi U^{\dagger}
$.
This action is also has ${\cal N}=2$ supersymmetry  
\[
\begin{array}{cc}
\left\{
\begin{array}{l}
\delta^{(1)} A_{\mu} =i \bar{\epsilon}\sigma_{\mu}\psi \\
\delta^{(1)} \psi  =\frac{i}{2}
([A_{\mu} ,A_{\nu} ] -i\alpha\epsilon_{\mu\nu\lambda}A_{\lambda})
\sigma^{\mu\nu}\epsilon
\end{array}
\right.
&
\mbox{, } 
\left\{
\begin{array}{l}
\delta A_{\mu}^{(2)} =0 \\
\delta ^{(2)}\psi  =\xi. 
\end{array} 
\right.
\end{array}
\]
The equations of motion corresponding to the action is 
\beqa
\left[A_\nu, \left[ A_\nu, A_\mu \right] + i \alpha
  \epsilon_{\mu\nu\lambda}A_\lambda \right]
&=&{1 \over 2}\left\{\psi_\beta, \psi_\alpha \right\}
(\sigma_0 \sigma_\mu)_{\alpha\beta} \cr
\left[\psi_\alpha, A_\mu \right](\sigma_0\sigma_\mu)_{\alpha\beta}
&=&0 ,\,\,\,\sigma_0=i \sigma_2. \label{eom1}
\eeqa
When $\psi = 0$,
the  typical solution for $A_\mu$ is, $A_\mu=X_\mu$, where
\bel{soln1}
[X_\mu, X_\nu ]=i \alpha \epsilon_{\mu\nu\lambda}(X_\lambda - R_\lambda),~~~~
[X_\mu, R_\nu]=0 
\eeq
represents an algebra of the fuzzy sphere configuration.
The commuting solution 
$X_\mu=$ diag$(r^{(1)}_\mu, r^{(2)}_\mu,r^{(3)}_\mu,\cdots,\cdots,r^{(N)}_\mu)$
is an special case of (\ref{soln1}). 

Remarkably, this equations of motion has solution which represent 
{\it arbitrary number of points or/and fuzzy spheres} 
with {\it various radius }and {\it centers}.
To see is, choose solution $X_\mu$ as block diagonal type,
\bel{fuz-model}
X_\mu
=
\left(
\begin{array}{cc}
\begin{array}{cc}
X^{(1)}_\mu &   \\
  &  X^{(2)}_\mu
\end{array} 
& 0 \\
0 &
\begin{array}{ccc}
X^{(3)}_\mu   & & \\
  & \ddots &  \\
   & &X^{(l)}_\mu   
\end{array}
\end{array}
\right)
\eeq
where $m$th  block $X^{(m)}_\mu$ is a $n_m \times  n_m$
irreducible representation of $SU(2)$
($\sum_{m=1}^l n_m = N$)
and obeys
\bel{fuzeqn1}
[X^{(m)}_\mu,X^{(m)}_\nu ]   =  i \a \e^{\mu \nu \l} (X^{(m)}_\l -R^{(m)}_\l) 
\eeq
where $R^{(m)}_\l$ is proportional to identity matrix $1_{n_m \times n_m}$.
\noindent
Though $Tr(X_\mu) = 0$, $X^{(m)}_\mu$ need not to be traceless.
These relation can be kept even when we assume relation, 
\bel{fuzeqn2}
\sum_{\l=1}^3 (X_\l^{(m)}-R_\l^{(m)})^2 = \rho_m^2 1_{n_m \times n_m}
\eeq
where $\rho_m^2= \a^2 \f{n_m^2-1}{4}$. Because of equation (\ref{fuzeqn1},
\ref{fuzeqn2}), we can think of this configuration is multi
fuzzy spheres. Now 
$r^{(m)}_\mu=\f{1}{n_m} Tr(X^{(m)}_\mu)=\f{1}{n_m} Tr(R^{(m)}_\mu) $ gives the co-ordinate of $m$th block
(center of $m$th fuzzy sphere or point) and $\rho_m$ is the radius. At this point, we can 
comment, the elements of the matrix $X_\mu$ and  the trace $r^{(m)}_\mu$
can be assumed to take any value from ${\bf R}$. Under this  
consideration  the space configuration is ${\bf R}^3$.\\

\noindent
{\bf Example 1. One Fuzzy Sphere Model}

To construct a one fuzzy sphere model out of this current 
scenereo, we assume $X_\mu$ to be of the form
\bel{onefuz}
X_\mu
=
\left(
\begin{array}{cc}
X^{(1)}_\mu & 0 \\
0 & 
\begin{array}{ccc}
r^{(2)}_\mu   & & \\
  & \ddots &  \\
   & &r^{(N-n_1+1)}_\mu 
\end{array}
\end{array}
\right)
\eeq
here, $n_{m>1} =1 $, $r^{(m>1)}_\mu \in {\bf R}$ and 
\beq
 [X^{(1)}_\mu,X^{(1)}_\nu ]  = i \a \e^{\mu \nu \l} (X^{(1)}_\l -R^{(1)}_\l).
\eeq 
This configuration represents 
a fuzzy sphere with center at 
$r^{(1)}_\mu = \f{1}{n_1} Tr(X^{(1)}_\mu)$
and ($N-n_1$) points at  co-ordinates
$r^{(m>1)}_\mu$ as back-ground. \\

\noindent
{\bf Example 2. Two Fuzzy Sphere Model}

We can construct  a multi-fuzzy sphere picture out of this
model. To construct a $k$ fuzzy sphere case, we 
consider the same configuration 
as equation(\ref{fuz-model}) with $k$ irreducible blocks
and $N-\sum_{m=1}^k n_m$ points. For example, for two fuzzy sphere
case we consider
\bel{twofuz}
X_\mu
=
\left(
\begin{array}{cc}
\begin{array}{cc}
X^{(1)}_\mu &   \\
  &  X^{(2)}_\mu
\end{array}
& 0 \\
0 &
\begin{array}{ccc}
r^{(3)}_\mu   & & \\
  & \ddots &  \\
   & &r^{(N-n_1-n_2+2)}_\mu
\end{array}
\end{array}
\right)
\eeq
here, $n_{m>2} =1 $, $r^{(m>2)}_\mu \in {\bf R}$ and 
\beq
 [X^{(1)}_\mu,X^{(1)}_\nu ]  = i \a \e^{\mu \nu \l} (X^{(1)}_\l -R^{(1)}_\l),~~
 [X^{(2)}_\mu,X^{(2)}_\nu ]  = i \a \e^{\mu \nu \l} (X^{(2)}_\l -R^{(2)}_\l)
\eeq 
This configuration represents 
two fuzzy spheres with centers at 
$r^{(1)}_\mu = \f{1}{n_1} Tr(X^{(1)}_\mu) =  \f{1}{n_1} Tr(R^{(1)}_\mu)$,
$r^{(2)}_\mu = \f{1}{n_2} Tr(X^{(2)}_\mu) =  \f{1}{n_2} Tr(R^{(2)}_\mu)$
and ($N-n_1-n_2$) points at  co-ordinates
$r^{(m>2)}_\mu$ as back-ground.

\subsection{One Loop Calculation for Two Blocks}

We assume one loop correction is good approximation for the interaction
between fuzzy spheres.

To see the effect of the fluctuation for this model, 
we expand the original matrices around these back ground
\beq
A_\mu=X_\mu + \tilde{A}_\mu, ~~~~~ 
\psi_\a=\chi_\a + \tilde{\vp}_\a ,\,\,\, \mbox{choose $\chi_\a =0$} 
\eeq
where $N\times N$ matrices $\tilde{A}$ and $\tilde{\vp}$ are 
quantum fluctuation.

\noindent
1-loop correction of effective action $W$ is calculated as
\beq
W= -ln \int d\tilde{A}\, d\tilde{\vp}\, e^{-S_2}
\eeq
where $S_2$ is quadratic terms of fluctuations in action 
(\ref{fuzaction}).
We add gauge fixing term and ghost term
\[
S_{gf} = -\f{1}{2 g^2} Tr [X_\mu, \tilde{A}_\mu]^2, ~~~~
S_{gh} = -\f{1}{g^2} Tr [X_\mu, B][X_\mu, C]
\]
 
\noindent
We re-write the $X_\mu$ in equation (\ref{fuz-model}),
as
\bel{fuzzyone}
X_\mu = 
\left( 
\begin{array}{cc}
Y^{(1)}_\mu & 0 \\
0 & Y^{(2)}_\mu
\end{array}
\right),~~
R_\mu = 
\left( 
\begin{array}{cc}
S^{(1)}_\mu & 0 \\
0 & S^{(2)}_\mu
\end{array}
\right)
\eeq
where, we treat $Y^{(1)}_\mu$ block as the fuzzy sphere(s)
and we regard $Y^{(2)}_\mu$ as background.

\noindent
We consider the fluctuations of the from
\[
\tilde{A}_\mu = \left(
\begin{array}{cc}
\tilde{A}_\mu^{(1)} & \tilde{B}_\mu \\
\tilde{B}_\mu^\dag & \tilde{A}_\mu^{(2)}
\end{array}
\right),~
\tilde{\vp}_a = \left(
\begin{array}{cc}
\tilde{\vp}_a^{(1)} & \tilde{\psi}_a \\
\tilde{\psi}_a^\dag & \tilde{\vp}_a^{(2)}
\end{array}
\right),~
B = \left(
\begin{array}{cc}
\tilde{B}_\mu^{(1)} & \tilde{D}_\mu \\
\tilde{D}_\mu^\dag & \tilde{B}_\mu^{(2)}
\end{array}
\right),~
C = \left(
\begin{array}{cc}
\tilde{C}_\mu^{(1)} & \tilde{E}_\mu \\
\tilde{E}_\mu^\dag & \tilde{C}_\mu^{(2)}
\end{array}
\right)
\] 
in terms of the above components, we can re-write
the bosonic, fermionic and
ghost parts of action (up to second order of the 
fluctuations) as 
 
\beqa
S_{2,B}
&=&{1 \over g^2} \sum_{p=1,2} Tr \left\{
-{1 \over 2} [\tilde{A}^{(p)}_\mu, Y^{(p)}_\nu]^2 
+ 2 i \a \e_{\mu \nu \l} S^{(p)}_\l \tilde{A}^{(p)}_\mu \tilde{A}^{(p)}_\nu
\right\} + \non
& & \!\!\!\!\! 
{1 \over g^2}
(\tilde{B}_\mu^\dag)_{Ii}
\!\! \left[({\cal H}^2) \delta_{\mu\nu} 
-2i \a \e_{\mu\nu\l}
\{(S^{(1)}_\l \otimes {\bf 1} + {\bf 1}\otimes S^{(2)}_\l)
+ {\cal H}_\l \}
\right]_{ijIJ}
\!\! (\tilde{B}_\nu)_{jJ}
\label{oneboson}  \\
S_{2,F}
&=& {1 \over 2 g^2}
\left\{\sum_{p=1,2}  Tr \tilde{\vp}^{(p)} \s^\mu [Y^{(p)}_\mu,
\tilde{\vp}^{(p)}] \right\} 
+{1 \over g^2}
\bar{(\tilde{\psi}^\dag)}_{Ii}
\left[
\s_\mu ({\cal H}_\mu)_{ijIJ}
\right]
(\tilde{\psi})_{jJ}
\label{onefermi} \\
S_{2,G} &=& {1 \over g^2}
\left\{\sum_{p=1,2}  Tr[Y^{(p)}_\mu, \tilde{B}^{(p)}]
[Y^{(p)}_\nu, \tilde{C}^{(p)}]
\right\}
\non & &
+{1 \over g^2}
\left\{
(\bar{\tilde{D}^\dag})_{Ii}
({\cal H}^2)_{ijIJ}
(\tilde{E})_{jJ}
-
(\bar{\tilde{E}^\dag})_{Ii}
(H^2)_{ijIJ}
(\tilde{D})_{jJ}
\right\}
\label{oneghost}
\eeqa
where       
\[
({\cal H}_\mu)_{ijIJ}=(Y_\mu^{(1)})_{ij}\otimes{\bf 1}_{IJ}- {\bf
  1}_{ij}\otimes{(Y_\mu^{(2)})_{IJ}}^*
\]
$i,j= 1\cdots dim(Y_\mu^{(1)})$,  $I,J= 1\cdots dim(Y_\mu^{(2)})$ and
"*" denote complex conjugate.

From this, we can get one-loop effective action for one fuzzy sphere 
or multi-fuzzy sphere system considering one block (irreducible)
or multi-block diagonal (reducible) form for $Y^{(1)}$.
In equation (\ref{oneboson}-\ref{oneghost}) each first term
represents self interaction of blocks of equation (\ref{fuzzyone})
and each second term represents interaction between two blocks. \\

\noindent
{\bf Example : One Fuzzy Sphere case}

For example, if we replace $Y^{(1)}$ by $X^{(1)}$  and 
$Y^{(2)}$ by rest diagonal part as equation
(\ref{onefuz}), we get the one loop effective action for one 
fuzzy sphere case. 
We can see, in such case
 the first part of individual equations gives the 
self-interaction term; $p=1$ term is  fuzzy sphere self 
interaction and $p=2$ term 
gives the self interaction between the 
space points. The second part
of the equations are the interaction between the fuzzy 
sphere and the space points. 

If, we take one fuzzy sphere at origin $ie$ $R_\mu^{(1)}=0$,
the self interaction term for one fuzzy sphere 
is
\beqa
S_{2,B}^{(self)}
&=&- {1 \over 2 g^2} Tr
[\tilde{A}^{(1)}_\mu, Y^{(1)}_\nu]^2 
\non
S_{2,F}^{(self)}
&=& {1 \over 2 g^2}
Tr  \tilde{\vp}^{(1)} \s^\mu [Y^{(1)}_\mu,
\tilde{\vp}^{(1)}]
\non
S_{2,G}^{(self)} &=& {1 \over g^2}
Tr[Y^{(1)}_\mu, \tilde{B}^{(1)}]
[Y^{(1)}_\nu, \tilde{C}^{(1)}]
\nn
\eeqa
we get similar expressions of self-interaction for 
the model in  \cite{iso1}.

Similarly, if we replace $Y^{(1)}$ by a block diagonal form with 
two irreducible blocks (as equation (\ref{twofuz}) ), we get 
the one loop effective action for two fuzzy sphere case. We 
will discussed about this in detail  in the following  section.

\section{One Loop Effective Action for Two Fuzzy Sphere system}

As we have seen in earlier section, we can calculate the one loop 
effective action from equations(\ref{oneboson}-\ref{oneghost}).
For this we consider the same configuration 
as equation(\ref{fuzzyone}), consider $Y^{(1)}_\mu$ 
as block-diagonal with two blocks (same as eqn. \ref{twofuz})
, each block representing one fuzzy 
sphere.  For two fuzzy
sphere configuration, we assume the following form 
for the back-ground  and fluctuation matrices.
\[
Y^{(1)}_\mu=\left( 
\begin{array}{cc}
X_\mu^{(1)} & 0\\
0 & X_\mu^{(2)}
\end{array}
\right),\,\,\,
\tilde{A}^{(1)}_\mu=\left(
\begin{array}{cc}
a_\mu^{(1)}   & b_\mu\\
b_\mu^\dag & a_\mu^{(2)}
\end{array}
\right),\,\,\,
\tilde{\vp}^{(1)}_\a
= \a s_\a,\,\,\,
s_\a=\left(
\begin{array}{cc}
s_\a^{(1)} & t_\a\\
t_\a^\dag & s_\a^{(2)}
\end{array}
\right)
\]
\bel{fluct}
\tilde{B}^{(1)}=\left(
\begin{array}{cc}
b^{(1)} & d\\
d^\dag & b^{(2)}
\end{array}
\right),\,\,\,
\tilde{C}^{(1)}=\left(
\begin{array}{cc}
c^{(1)} & e\\
e^\dag & c^{(2)}
\end{array}
\right),~
S_\l^{(1)} = 
\left(
\begin{array}{cc}
R_\l^{(1)} & 0\\
0 & R_\l^{(2)}
\end{array}
\right)
\eeq
where, in these matrices,
the first diagonal block is $n_1 \times n_1$ matrix and
the second is $n_2 \times n_2$.  
For calculational simplicity, we further assume 
$Y^{(2)}_\mu=0, R_\l^{(2)}=0$.
$R_\l^{(i=1,2)} = r_\l^{(i)} {\bf 1}_{n_i \times n_i}$.
$r_\l^{(1)}= {n_2 \over n_1+n_2} r_\l$,  
$r_\l^{(2)}=- {n_1 \over n_1+n_2} r_\l$, 
are the centers of two fuzzy spheres and $r_\l$ is the 
distance vector between them. 
\[
\tilde{D}=
\left( 
\begin{array}{c}
D^{(1)} \\ D^{(2)}
\end{array}
\right),~
\tilde{E}=
\left( 
\begin{array}{c}
E^{(1)} \\ E^{(2)}
\end{array}
\right),~
\tilde{B_\mu}=
\left( 
\begin{array}{c}
B^{(1)}_\mu \\ B^{(2)}_\mu
\end{array}
\right),~
\tilde{\psi_a}=
\left( 
\begin{array}{c}
\psi^{(1)}_a \\ \psi^{(2)}_a
\end{array}
\right)
\]
where, in these matrices,
the upper  block is $n_1 \times (N-n_1-n_2)$ matrix and
the second is $n_2 \times (N-n_1-n_2)$.

\noindent
Putting these in equation (\ref{oneboson} - \ref{oneghost}),
the total bilinear terms can be written as the sum of 
following 9 terms. 

\beqa
S_{2,B}^{(self)} &=&{1 \over g^2} \sum_{i=1,2} Tr \left\{
-{1 \over 2} [a^{(i)}_\mu, X^{(i)}_\nu]^2 
+ 2 i \e_{\mu \nu \l} R^{(i)}_\l a^{(i)}_\mu a^{(i)}_\nu
\right\} \label{eqn1}\\  
S_{2,F}^{(self)}&=& {\a^2\over 2 g^2}
\sum_{i=1,2} Tr \left\{s^{(i)} \s^\mu [L^{(i)}_\mu,
s^{(i)}] \right\} 
\\  
S_{2,G}^{(self)}&=& 
{\a^2 \over g^2}
\left\{ \sum_{i=1,2} Tr[L^{(i)}_\mu, b^{(i)}]
[L^{(i)}_\nu, c^{(i)}]
\right\}
\\  
S_{2,B}^{(back)} &=&
{\a^2 \over g^2}
\sum_{i=1,2}
(\tilde{B}^{(i)^\dag}_\mu)_{Ik_i }
\left[(L^{(1)}_\rho \otimes {\bf 1})^2 \delta_{\mu\nu}
-2i \e_{\mu\nu\l} c^{(i)}_\l {\bf 1}
\right]_{k_i l_i IJ}
(\tilde{B}^{(i)}_\nu)_{l_i J}
\\  
S_{2,F}^{(back)}
&=&
{\a^2 \over 2 g^2}
\sum_{i=1,2}
\bar{(\tilde{T})}_{Ik_i}
\left[
\s_\mu (L^{(i)}_\mu)_{k_i l_i IJ}
\right]
(\tilde{T})_{l_i J}
\\  
S_{2,G}^{(back)} &=& {\a^2  \over g^2}
\sum_{i=1,2}
\left\{
(\bar{\tilde{D}^\dag})_{Ik_i }
(L^{(i)} \otimes {\bf 1})^2_{k_i l_i IJ}
(\tilde{E})_{l_i J}
-
(\bar{\tilde{E}^\dag})_{Ik_i }
(L^{(i)} \otimes {\bf 1})^2_{k_i l_i IJ}
(\tilde{D})_{l_i J}
\right\} 
\\  
S_{2,B}^{(1)(2)} &=& 
{\a^2 \over g^2}
(b_\mu^\dag)_{k_2 k_1 }
\left[(H^2) \delta_{\mu\nu}
-2i \e_{\mu\nu\l}
c_\l \otimes {\bf 1}
\right]_{k_1 l_1 k_2 l_2 }
(b_\nu)_{l_1 l_2}
\\  
S_{2,F}^{(1)(2)}
&=& 
{2 \a^2 \over g^2}
(\bar{t})_{k_2 k_1}
\left[
\s_\mu ( H_\mu)_{k_1 l_1 k_2 l_2 }
\right]
(t)_{l_1 l_2}
\\  
S_{2,G}^{(1)(2)} &=& 
{\a^2  \over g^2}
\left\{
(\bar{d})_{k_2 k_1 }
(H^2)_{k_1 l_1 k_2 l_2 }
(e)_{l_1 l_2 }
-
(\bar{e})_{k_2 k_1 }
(H^2)_{k_1 l_1 k_2 l_2 }
(d)_{l_1 l_2 }
\right\} 
\label{eqn9}
\eeqa
where       
\[
(H_\mu)_{k_1 l_1 k_2 l_2 }=(L_\mu^{(1)})_{k_1 l_1 }\otimes{\bf 1}_{k_2 l_2 }- {\bf
  1}_{k_1 l_1 }\otimes{(L_\mu^{(2)})_{k_2 l_2 }}^*
\]
$k_i,l_i= 1\cdots n_i$,  $I,J= 1\cdots (N-n_1-n_2)$,
$L^{(i)}_\mu={1 \over \alpha}X^{(i)}_\mu,~
c_\mu={1 \over \alpha}r_\mu.
$

\noindent
We can see each of bosonic, fermionic and ghost has three 
parts describing self-interaction (denoted by $S_2^{(self)}$),
interaction between two fuzzy spheres (denoted by $S_2^{(1)(2)}$) 
and the extra piece
coming from the interaction of each fuzzy sphere
with the back ground (denoted by $S_2^{(back)}$ index $i$, for i-th one). We can as
well say these extra piece as part of the self energy of 
the fuzzy spheres because they exist even for one fuzzy sphere 
case (section 2.2).

\subsection{Interaction Between Two  Fuzzy Spheres}

We assume one loop correction is good approximation for the interaction
between fuzzy spheres \footnote{This is not a good approximation for two 
intersecting fuzzy spheres, for  $eg$, $N_1 = N_2, c=0$.}.
1-loop correction of effective action $W$ is calculated as
\beq
W= -ln \int da\, ds\, db\, dc\, e^{-S_2}
\eeq
As the the total action $S$ decouples into each sector, we can write  
\beq 
W= W^{(self)}_{(B+F+G)}+W^{(back)}_{(B+F+G)}+W^{(1)(2)}_B+W^{(1)(2)}_F+W^{(1)(2)}_G
\eeq
where indices correspond to those of 
equations(\ref{eqn1}-\ref{eqn9}).
 
\noindent 
We are now interested in following parts
those are from interactions between two  fuzzy spheres.
\[
\begin{array}{ll}
W^{(1)(2)}_B&
=-ln \int db\,db^\dag
e^{-b_\mu^\dag
\left[H^2 \delta_{\mu\nu} -2i \e_{\mu\nu\l}c_\l
\right]
b_\nu}\\
&=-ln\left[det^{-{1 \over 2}}\left(H^2 - 2i\e \cdot c \right)\right]^2
\\
W^{(1)(2)}_F&
=-ln\int dt\,dt^\dag
e^{-\bar{t^\dag}
\left[ \s_\mu H_\mu \right]
t}\\
&=-ln\left[ det^{1 \over 2} \s\cdot H\right]^2
 \\                              
W^{(1)(2)}_G&
=-ln\int dd\,dd^\dag\,dc\,dc^\dag
e^{-d^\dag H^2 e +e^\dag H^2 d}
\\
&=-ln\left[ det H^2\right]^2
 \\
\end{array}
\]
where
squares of determinants come from two off-diagonal blocks of matrices
and ${1 \over 2}$ in $W^{(1)(2)}_F$ is because of Majorana spinor.
        
\subsubsection{Bosonic Sector}

Without loss of generality,
two fuzzy spheres are assumed to be separated by $r$ in 3rd direction
\[
c_\mu=(0,0,c),~~ r=\a c
\]
Then diagonalise the operator in bosonic part
\[
H^2-2i\e\cdot c=\left(
\begin{array}{ccc}
H^2 & -2ic & 0\\
2ic & H^2 &0\\
0 & 0& H^2
\end{array}
\right)
\sim\left(
\begin{array}{ccc}
H^2-2c &0&0\\
0& H^2+2c &0\\
0&0& H^2
\end{array}
\right)
\]
So, we can write the bosonic contribution to the effective 
action (including the ghost part) as,

\beq
W_B^{(1)(2)}=
{1 \over 2} \mbox{Ln} \left[ \det \left(
\frac{\left(H^2-2c\right)\left(H^2+2c\right)}
{H^2}
\right) \right]
\eeq

\noindent
We define
$ J_\mu = H_\mu + c_\mu$ and
$K_\mu = J_\mu + S_\mu$,
where $S_\mu = {\s_\mu \over 2}$.
Both $J_\mu$ and $K_\mu$ follow $SU(2)$ algebra.
$j$, maximum eigen value of $J_3$,   varies from $j_{min}=|\f{n_1-n_2}{2}|$ to $j_{max}=(\f{n_1+n_2}{2}-1)$ and
$k=j\underline{+} \f{1}{2}$. $H^2, H^2\underline{+}c$
or $H\cdot\s$) are block diagonal,
each block representing a particular value of $j$.
So, we can write
\bel{onel2}
W^{(1)(2)}_{(B+G)}={1 \over 2} ln \left[
\prod_{j=j_{min}}^{j_{max}}
(w_B)_j \right]
\eeq
where, $(w_B)_j$ is the determinant of $j$th block and can be
calculated to be 
\beq
(w_B)_j = \f{(c^2 - 2 c (j+1) +j(j+1))^2}{(c^2 + 2 c j+j(j+1))^2}
\prod_{i=-j}^j {(c^2 + 2 c (-i+1) +j(j+1))^2}
\eeq

\noindent
For $c << 1$, when the fuzzy spheres are co-centric, there is a non-zero interaction 

\beqa
W^{(1)(2)}_{(B+G)} & =& \sum_{j=j_{min}}^{j_{max}}
(2j+1) ln[j(j+1)] 
\non
&-& \! \f{2}{3} 
\left[\! 
\f{1}{n_1+n_2}\!\! \left(1 + \f{24}{n_1+n_2} \right) 
-\f{1}{n_1-n_2}\!\! \left(1 + \f{24}{n_1-n_2} \right) 
+ \sum_{j=j_{min}}^{j_{max}} \f{1}{j}
\right]
c^2 
+ O\left(c^4\right). \nn
\eeqa

\noindent
For $c >> 1$ $ie$ when the fuzzy spheres are far apart, 
\bel{largecboson}
W^{(1)(2)}_{(B+G)} = n_1 n_2 \log c^2 + \f{n_1 n_2}{4} (n_1^2 +n_2^2 -18)\f{1}{c^2}
+ O\left(\f{1}{c^4}\right).
\eeq

\subsubsection{Supersymmetric Case }

Summing up all these contributions, 
\beq
W^{(1)(2)}=
{1 \over 2} ln \left[ \det \left( 
\frac{\left(H^2-2c\right)\left(H^2+2c\right)}
{H^2\left( \s\cdot H \right)^2}
\right) \right]
\eeq

\noindent 
We can write
\bel{onel1}
W^{(1)(2)}={1 \over 2} ln \left[ 
\prod_{j=j_{min}}^{j_{max}}
w_j \right]
\eeq
where, $w_j = \tilde{w_j}(c) \tilde{w_j}(-c)$ 
is the determinant of $j$th block and

\beqa
\tilde{w_j}(c) =
\f{\left[c^2 +c (2 j+1) + j(j+1) \right]
\left[c^2 +2 c (j+1) + j(j+1) \right]}
{(c+j)^2 
\left[ c^2 +2 c j + j(j+1) \right]} ~~~~\non
\times
\prod_{i=-j}^{j} 
\f{\left[c^2 +2 c (i+1) + j(j+1) \right]}
{\left[c^2 + c (2i+1) + j(j+1) \right]}.
\label{onel2}
\eeqa

When $c=\infty$, $ie$ the fuzzy spheres are at large distance,
$W^{(1)(2)} =0 $,  $ie$
two fuzzy spheres do not interact
each other when they are far apart.
This feature is different from bosonic case. This is because of  
some cancelation between bosonic and the fermionic contributions.

Expanding $W^{(1)(2)}$ around $c=0$ and $c=\infty$, 
we can get the potential between two fuzzy sphere for small and 
large distance. 

\noindent 
For small $c$,

\bel{potn0}
W^{(1)(2)}_{c << 1}
=
\log \left| \f{n_1+n_2}{n_1-n_2} \right| 
- 96 \f{n_1 n_2}{(n_1+n_2)^2(n_1-n_2)^2} c^2
+O\left(c^4\right).
\eeq

\noindent
We see an attractive force between the fuzzy sphere surfaces
(repulsive between the centers), when the centers are close to each
other. 

\noindent
For large $c$, 

\bel{potni}
W^{(1)(2)}_{c >> 1}
=
- 8 n_1 n_2 / c^2 + O\left(\f{1}{c^4}\right).
\eeq

So, there also exists an attractive force between the fuzzy spheres when they are 
at large distances. This result is consistent with the earlier observation
that D2(D0)-branes form a bigger D2-brane in  string theory \cite{my,hikida}.
We can compare this to equation (\ref{largecboson}) and find some cancelation
between bosonic and fermionic part, that is because of part of supersymmetry.

Spherical BPS D2-brane from string theory in $S^2$ does not interact
when they are "parallel" ($c=0$) to each other. However from our approach 
in matrix model the first term of equation (\ref{potn0}) exists 
 even for 
$c=0$. It may be explained as the quadratic part of action $S^2$ (eqn. 3.7-3.9)
itself is 
not supersymmetric under the susy transformations in eqn( 2.2).
But for large distance case $(c \rightarrow \infty)$ 
this $S^2$ recovers ${\cal N} =1$ supersymmetry and in such case 
$W^{(1)(2)}$ vanishes. Moreover in such case, we can not use only quadratic 
part of off-diagonal part in equation (\ref{fluct}) when  the size of 
$n_1$ and $n_2$ is not so different, that corresponds to overlapping of
surface of two spheres.

\section{Conclusion}

In this paper, we have presented  a general fuzzy sphere model in three
dimension, which allows
multi fuzzy sphere system with discretely arbitrary radii and 
arbitrary location in ${\bf R}^3$.  We have added a Chern Simon term 
to the reduced model of 3D SYM. In original  model 
the space and  branes ($eg$ fuzzy spheres) are not separately 
distinguishable.  We have  artificially partitioned  the matrices
into multiple block diagonal form. In such case, the classical solution 
represents a system of space and fuzzy spheres (branes).
Classically these fuzzy spheres  and space are
non-interacting. We have tried to calculate the interaction
as the one loop quantum effect. 
In section 2 we have studied the fuzzy 
sphere and space time. We have calculated interaction of 
fuzzy spheres and space (\ref{oneboson} - \ref{oneghost}).
In section 3, one loop interaction of fuzzy spheres in 
bosonic and supersymmetric case are studied. 
In particular, we have calculated the interaction between 
two fuzzy spheres with radii  ($\rho_1 \sim \a n_1,
\rho_2 \sim \a n_2$)  ($n_1$ and $n_2$ is arbitrary)  at distance ($r= \a c$).
We have determined the one loop effective action for such system for 
both bosonic case and supersymmetric case for two concentric 
fuzzy spheres ($c << 1$) and for two fuzzy spheres
kept far apart ($c >> 1$).
There is a cancelation between bosonic and
fermionic part. In supersymmetric case, there is an attractive 
force between two fuzzy sphere surfaces for both large and small 
distance case. Even this model has a ${\cal N} = 2$ supersymmetry,
the one loop contribution for concentric case is 
non-zero for nearly equal $n_1$ and $n_2$. This is probably
because of the fact the one loop approximation is not good 
approximation in such case. It will be interesting to compare 
equations (\ref{potn0}, \ref{potni}) with those from spherical
D2-brane interactions in $SU(2)$ WZW model.

\begin{center}
{\bf Acknowledgments}
\end{center}

We would like to thank T. R. Govindarajan, N. D. Hari Dass, 
T. Jayaraman  and B. Sathiapalan for useful discussions 
and suggestions. 
H. T. would like to thank A. Sen for fruitful discussions 
during  his visit in HRI.

\end{document}